\documentclass[sigconf,nonacm]{acmart}
\usepackage{booktabs}
\usepackage{graphicx}
\usepackage{amsmath}
\usepackage{caption}
\usepackage{threeparttable}
\title{The Hidden Costs of Gatekeeping in Open Science: Quasi-Experimental Evidence from arXiv's Endorsement Policy Tightening}

\author{Wenchang Wang}
\affiliation{%
  \institution{Northwestern University}
  \department{Department of Communication Studies}
  \city{Evanston}
  \country{United States}
}
\email{wenchangwang2031@u.northwestern.edu}

\begin{document}

\begin{abstract}
Preprint servers and open science platforms have become core infrastructure through which scientific knowledge circulates on the Web, greatly facilitating the dissemination and diffusion of online scientific information. Yet to safeguard content quality, many of these platforms rely on existing members to vouch for newcomers; how such network-mediated gatekeeping reshapes entry barriers across different scientific communities remains systematically underexplored. To address this, I examine the potential unequal impacts of this mechanism through arXiv's tightening of its endorsement policy in early 2026. Using the metadata of three million arXiv papers and the co-authorship networks of each research community, I compare the influx and dynamics of first-time submitters before and after the policy across eight disciplines. The findings show that the policy reduced the entry of new authors, and this reduction was concentrated in fields where co-authorship ties are sparse, while fields with dense collaboration networks were largely spared. A comparison with a parallel preprint server in biology suggests that this likely reflects a shift in where researchers choose to post rather than a decline in research activity. These findings demonstrate that the network structure of a scientific community shapes who can enter online scientific production, advancing our understanding of how the Web's open knowledge infrastructure shapes inequality in the diffusion of science.
\end{abstract}

% ACM Computing Classification System (CCS) Concepts


\keywords{open science, preprints, gatekeeping, endorsement systems, academic inequality, quasi-experiment, co-authorship networks}

\maketitle

%==============================================================================
\section{Introduction}
%==============================================================================

Preprint servers have become the primary route through which new scientific findings reach readers on the Web~\cite{ginsparg2011arxiv}. These servers (such as arXiv, bioRxiv, and SSRN) have grown into vital repositories for scientific advancement and research, allowing researchers to share cutting-edge findings without waiting for peer review~\cite{suber2012open, sever2019biorxiv}. As scientific diffusion becomes deeply intertwined with the Web, these platforms matter not only for which scientific outputs are read, developed, and built upon, but also for who is able to participate in scientific research and gain visibility~\cite{xie2021preprint}. Prior work widely credits these platforms with lowering the cost of participating in scholarly communication, thereby weakening the inherent advantages enjoyed by researchers at well-resourced institutions—such as greater visibility in peer review and the influence of prestige on scientific production~\cite{bjork2010open, xie2021preprint}. However, openness and complete equality on the Web are rarely unconditional~\cite{bjork2010open}; moreover, in scientific research, unconditional openness and equality can introduce potential issues such as declining content quality~\cite{xie2021preprint}. For a long time, driven by content moderation and management needs, these open platforms have implemented gatekeeping policies, such as requiring institutional affiliations or institutional email addresses to verify the identity of paper submitters. Yet what consequences arise when such requirements are escalated?

Research on inequality in science has long shown that access to scholarly communities is often built upon social ties. Established scientists leverage these social relationships and capital in ways that crowd out newcomers~\cite{merton1968, way2019productivity}. Collaboration networks serve as a key conduit and manifestation of such social ties and resource acquisition, shaping who can learn community norms and find co-authors more quickly and effectively~\cite{newman2001, newman2004coauthorship, dong2017century}. Furthermore, social networks are widely recognized as fueling the Matthew Effect in scientific production, as advantages and disadvantages tied to position, status, and background accumulate across an entire career~\cite{ginther2011race, clauset2015systematic, teplitskiy2022status}. However, past research on how social networks generate inequality in scientific participation and production mostly focuses on outcomes after entering academia (such as publications, citations, and funding acquisition), taking the ``opportunity to participate'' itself as a given. Much less attention has been paid to whether the entry point itself—the exact moment a researcher first gains access to a publication venue—depends on the network structure of their community. Meanwhile, research on knowledge production and peer collaboration indicates that although entry rules are often intended to curb low-quality submissions, establishing such gatekeeping mechanisms also deters high-quality contributions~\cite{hill2020hidden}; more importantly, entry gatekeeping creates disproportionate impacts across different groups~\cite{wagner2015man, patro2020fairrec}. Yet these studies primarily examine general-purpose platforms such as encyclopedias and social media, where entry rules are treated purely as platform-level attributes, leaving a gap when it comes to genuine scientific participation.

This leaves a research gap at the intersection of these two literatures: we lack empirical evidence on how such ``network-mediated gatekeeping'' at the entry point of open science platforms affects participation, and whether its burden is distributed evenly across scientific communities or concentrated in fields where collaborative ties are scarce. This question is critical because the cost imposed by a personal endorsement requirement depends on a newcomer's social capital~\cite{granovetter1973}; in other words, from the very moment of entry, newcomers must bear the inequalities and corresponding consequences dictated by their social capital~\cite{backstrom2006group}.

To address this gap, I treat arXiv's tightening of its endorsement policy in 2026 as a natural experiment. Following an earlier rollout in mathematics in December 2025, arXiv further narrowed the scope in January 2026 under which an institutional email address alone suffices for approval, forcing more newcomers to find an existing author willing to vouch for them. This constitutes an exogenous shock to the entry rules of scientific production. To this end, I construct a panel dataset of arXiv subcategories tracking the influx of first-time submitters over time, build a pre-policy co-authorship network for each subcategory, and estimate the effect of the policy on new-author entry using event-study, difference-in-differences (DID), and regression discontinuity designs (RDD). Next, I examine how this effect varies with the density of a community's co-authorship network and compare arXiv with bioRxiv, a parallel preprint server unaffected by the policy. The findings reveal that the policy reduced the overall influx of new authors by approximately 13\%. This decline was concentrated in disciplines with sparse co-authorship networks (such as economics and quantitative biology), where nearly half of the expected new entrants failed to appear, whereas densely connected fields such as physics were largely unaffected.

The contributions of this paper are threefold. Methodologically, I provide evidence on how entry barriers on scholarly Web platforms alter participation in scientific production. Theoretically, my research highlights the unintended consequences of online governance, demonstrating that this effect is not uniform but instead disproportionately impacts disciplines with sparse versus dense networks. Practically, this study offers design implications for open knowledge platforms, exploring how such platforms can maintain quality control without making access overly dependent on pre-existing social ties.

\section{Related Work and Hypotheses}
%==============================================================================

\textbf{Barriers to Entry in Online Knowledge Production}

In web-mediated knowledge production systems, lowering participation costs is widely regarded as the primary driver of large-scale open contribution~\cite{benkler2006, bennett2012}. Such low thresholds facilitate the entry of newcomers, who serve as a vital resource and driving force for the operation and sustained growth of these ecosystems~\cite{wang2026}. However, although classical theories view barriers to entry as costly constraints that deter participation~\cite{olson1965, fulk1996}, platforms routinely deploy gatekeeping mechanisms—such as identity verification, credentialing, and content moderation—to safeguard against spam, vandalism, and low-quality submissions, thereby ensuring the smooth functioning of the system~\cite{kumar2015analysis, nilizadeh2019think, jhaver2019transparency, ribeiro2021platform}. This practice reflects a core governance tension between quality control and inclusivity~\cite{kraut2012building, ostrom1990}.

Empirical studies in Web science demonstrate that the design of these entry rules fundamentally shapes both the identity of contributors and the quality of their contributions~\cite{halfaker2013rise, kraut2012building, wang2026}. Even minimal procedural requirements—such as creating an account, passing a CAPTCHA test, or obtaining an institutional credential—create friction and participation costs for newcomers~\cite{kollock1999, halfaker2011dont, kooti2012predicting, kelty2008two, hill2020hidden}. Crucially, such barriers rarely function as precise filters for quality.

In the context of scholarly communication, preprint servers were originally designed to lower participation costs and democratize access to the scientific record~\cite{bjork2010open, xie2021preprint}. Yet arXiv's endorsement requirement reintroduces a stringent social barrier at the point of entry: new authors who cannot rely on recognized institutional email domains must actively locate and convince an existing, qualified author to vouch for their work. Unlike the minor administrative friction of registering an account or verifying an email address, securing a personal endorsement entails substantial relational and psychological costs. It requires submitters to possess pre-existing social capital—specifically, a direct or indirect connection to an established insider embedded in the target subject category~\cite{lin2001}—and a willingness to expend personal favors and reputation before a paper is even made public. This requirement clearly exemplifies the entry measures discussed above that increase participation friction and costs. Accordingly, I hypothesize:

\medskip
\noindent \textbf{H1:} \textit{Tightening arXiv's endorsement policy will reduce the rate of new submitter entry.}
\medskip

\textbf{Network-Dependent Gatekeeping and the Matthew Effect}

Although platform governance research typically treats entry rules as uniform properties of the platform itself, a parallel literature on social capital and academic stratification suggests that barriers mediated by social networks do not bind all communities equally~\cite{merton1968, lin2001, clauset2015systematic}. \citet{merton1968} introduced the famous ``Matthew Effect,'' which in scientific knowledge production manifests as established scientists accumulating disproportionate recognition and resources by crowding out newcomers~\cite{way2019productivity}. Meanwhile, disadvantages tied to institutional prestige, status, and demographic background accumulate across an entire academic career~\cite{ginther2011race, clauset2015systematic, teplitskiy2022status}, generating further inequality.

In web-based social systems, the ability to acquire social capital is fundamentally governed by an individual's position within the underlying network topology (which reflects real-world interpersonal relationships~\cite{romero2011differences}). Because network topologies across different topical communities or domains exhibit distinct structural properties and information diffusion mechanisms~\cite{granovetter1973, backstrom2006group, leskovec2010signed, romero2011differences, alshebli2018preeminence}, the process of acquiring social capital can entail vastly different levels of difficulty and patterns across individual subfields. Specifically in scientific knowledge production, co-authorship networks constitute the core social infrastructure~\cite{newman2001, newman2004coauthorship, barabasi2002, dong2017century}, and scientific disciplines vary dramatically in their collaborative norms and resulting network topologies~\cite{newman2001, lariviere2013bibliometrics}.

Consequently, when an open platform conditions entry on peer endorsement, it effectively delegates gatekeeping power to these underlying collaboration networks. For an entry system predicated on social capital recognition, the effective cost of entry is therefore inversely proportional to the density of collaborative ties in a given field~\cite{backstrom2011supervised, crandall2008feedback}. In densely connected disciplines, the co-authorship network exhibits a giant connected component with short path lengths; any newcomer is likely embedded in a laboratory or team within one degree of separation from multiple eligible endorsers~\cite{lariviere2013bibliometrics}. In these fields, the cost of securing an endorsement is logically far lower than in sparsely connected disciplines where collaborative ties are scarce. In the latter, the exact same platform rule becomes an insurmountable barrier because potential endorsers are socially distant, forcing newcomers to bridge structural holes or solicit strangers to gain access~\cite{granovetter1973}. Therefore, I hypothesize:

\medskip
\noindent \textbf{H2:} \textit{The negative effect of endorsement tightening on new-author entry will be moderated by pre-policy co-authorship network density, such that disciplines with sparser co-authorship networks experience larger declines in new-author entry.}
\medskip

\textbf{Exclusion as Structural Inequality}

If the burden imposed by network-mediated gatekeeping depends on a community's topological structure, then tightening endorsement rules does not merely reduce participation in aggregate—it systematically redistributes access opportunities across the entire scientific landscape. Research on fairness and bias in online platforms indicates that seemingly neutral algorithmic or governance rules often produce highly uneven participation outcomes across different user groups~\cite{wagner2015man, patro2020fairrec}. In the context of open science infrastructure, applying a uniform ``acquaintance vouching'' requirement across structurally disparate disciplines effectively institutionalizes a form of structural inequality within the platform architecture~\cite{way2019productivity, lariviere2013bibliometrics}.

Preprint servers are widely credited with leveling the academic playing field by mitigating the institutional prestige bias inherent in traditional peer review~\cite{bjork2010open, xie2021preprint, teplitskiy2022status}. Yet by tethering submission rights to pre-existing co-authorship networks, strict endorsement policies subvert this democratizing promise. In fields at the very bottom of the network density spectrum, fragmented social structures offer almost no organic pathways to established insiders; here, the policy ceases to be a mild quality filter and instead evolves into an exclusionary boundary~\cite{patro2020fairrec}. Researchers in peripheral or emerging subfields, independent or interdisciplinary scholars entering a new domain without local ties, and newcomers from non-core institutions all face severe structural disadvantages. Rather than screening submissions based on scholarly merit, the platform screens entrants based on pre-existing social connectivity, which risks completely cutting off the inflow of new contributors in sparsely networked communities. Accordingly, I hypothesize:

\medskip
\noindent \textbf{H3:} \textit{Fields with the sparsest co-authorship networks will experience the largest absolute declines in new-author entry, resulting in the near-exclusion of newcomers.}

%==============================================================================
\section{Data and Measurement}
%==============================================================================

\subsection{Quasi-Experimental Setting and Data Collection}

arXiv is one of the most prominent online scientific platforms, spanning a wide array of disciplines including computer science, mathematics, physics, and economics~\cite{ginsparg2011arxiv}. For a long time, arXiv has required new submitters to obtain an ``endorsement'' before submitting and publishing papers; however, across most disciplines, possessing an institutional email address was sufficient to receive this endorsement automatically. In response to a surge in low-quality, non-scientific submissions, arXiv further raised this participation threshold. Specifically, on December 10, 2025, arXiv first tightened this rule in the Mathematics section: to qualify for automatic endorsement, submitters were not only required to provide an institutional email address, but also had to have previously authored a paper within that section; all other first-time submitters were required to secure a personal endorsement from an established author~\cite{boboris2025endorsementmath}. On January 21, 2026, this policy was extended to all subject sections on arXiv~\cite{boboris2026endorsement}. This staggered rollout strategy provides a quasi-experimental setting for this study to examine the potential impacts of participation barriers.

Through the arXiv API, I obtained the metadata snapshot comprising 3,021,763 papers and 625,715 unique first-time submitters. Based on this dataset, I constructed a weekly panel dataset covering 143 subcategories over 94 weeks (July 1, 2024 to April 14, 2026), yielding 10,754 subcategory-week observations that meet the baseline activity threshold ($\geq 5$ total weekly submissions).

My primary dependent variable is the count of \textit{new submitters} per subcategory-week---defined as authors uploading a paper to arXiv for the first time in the platform's entire history. Total weekly submissions per subcategory serve as the exposure variable to control for baseline fluctuations in overall submission volume.

\subsection{Co-authorship Network Construction}

To prevent post-treatment behavioral contamination (i.e., endogenous network adaptation), I construct a static co-authorship network for each subcategory using exclusively pre-treatment data (July 1, 2024--December 9, 2025)~\cite{newman2001}. For each paper with $k \geq 2$ authors, I project $\binom{k}{2}$ undirected edges connecting all co-author pairs. My primary network metric is \textbf{average degree}:
\begin{equation}
\bar{d}_c = \frac{2 |E_c|}{|V_c|}
\end{equation}
where $|E_c|$ denotes the number of unique co-authorship edges and $|V_c|$ denotes the number of unique authors in subcategory $c$. Average degree captures the typical number of distinct collaborators per researcher, serving as a direct proxy for how readily a newcomer can locate an eligible endorser. In regression models, I apply a log transformation, $\log(\bar{d}_c + 1)$, to accommodate the substantial right-skewness of network degree across fields.

\subsection{Treatment Definition and Summary Statistics}

The binary treatment indicator $\text{Post}_{ct}$ equals one for weeks following the enforcement of the tightened endorsement policy: December 10, 2025 for mathematics subcategories (which underwent an early preparatory rollout) and January 21, 2026 for all remaining subcategories. Table~\ref{tab:desc} presents summary statistics for the analytical sample. Pre-treatment average degree exhibits pronounced heterogeneity across disciplines (Kruskal-Wallis $\chi^2 = 81.61$, $p < 0.001$), spanning from $1.28$ in specialized theoretical subfields to over $930$ in large-scale experimental physics collaborations.

\begin{table*}[t]
\centering
\begin{threeparttable}
\caption{Descriptive Statistics for the Subcategory-Week Panel}
\label{tab:desc}
\begin{tabular}{lccccccc}
\toprule
Variable & Mean & SD & P25 & P50 & P75 & Min & Max \\
\midrule
New Submitters ($Y_{ct}$) & 9.71 & 18.93 & 2 & 4 & 10 & 0 & 202 \\
Total Submissions ($\text{Total}_{ct}$) & 46.23 & 79.96 & 11 & 22 & 52 & 5 & 987 \\
New Submitter Rate ($Y_{ct} / \text{Total}_{ct}$) & 0.208 & 0.127 & 0.124 & 0.200 & 0.280 & 0 & 1.000 \\
Pre-treatment Avg Degree ($\bar{d}_c$) & 35.32 & 103.29 & 3.14 & 6.19 & 12.00 & 1.28 & 930.81 \\
\bottomrule
\end{tabular}
\begin{tablenotes}
\small
\item \textit{Notes:} $N = 10{,}754$ subcategory-week observations across 143 subcategories and 94 weeks. A new submitter is defined as a first-time uploader in arXiv's complete publication history. Pre-treatment average degree is computed over the baseline period prior to any policy rollout (July 1, 2024 to December 9, 2025).
\end{tablenotes}
\end{threeparttable}
\end{table*}

%==============================================================================
\section{Analytic Strategy}
%==============================================================================

\subsection{Main Specification}

Because new submitter counts are non-negative integers exhibiting severe overdispersion ($\text{Var}/\text{Mean} = 36.9$; Cameron-Trivedi test $z = 11.64$, $p < 0.001$), ordinary least squares (OLS) models with log-transformed outcomes yield inconsistent estimates in the presence of heteroskedasticity. Following \citet{wooldridge2010} and \citet{cameron2013}, I estimate Poisson quasi-maximum likelihood (QMLE) models with two-way fixed effects:
\begin{equation}
\ln E[Y_{ct}] = \beta \cdot \text{Post}_{ct} + \alpha_c + \lambda_t + \ln(\text{Total}_{ct})
\label{eq:main}
\end{equation}
where $Y_{ct}$ is the count of new submitters in subcategory $c$ during week $t$; $\alpha_c$ denotes subcategory fixed effects, which absorb time-invariant disciplinary scale and submission norms; $\lambda_t$ represents year-month fixed effects capturing common macroeconomic and academic calendar shocks; and $\ln(\text{Total}_{ct})$ enters with its coefficient constrained to unity as an exposure offset. 

Under the Poisson QMLE framework, the estimator $\hat\beta$ remains consistent as long as the conditional mean is correctly specified, regardless of arbitrary overdispersion~\cite{hausman1984econometric, wooldridge2010}. The coefficient $\beta$ directly identifies the semi-elasticity (proportional change) in new submitter entry: $\Delta\% = e^{\beta} - 1$. Because the endorsement policy is administered at the broader endorsement-domain level (8 major disciplines), I cluster standard errors at the endorsement-domain level and apply small-sample $t(G-1) = t(7)$ degrees-of-freedom adjustments to prevent over-rejection~\cite{cameron2008}.

\subsection{Network Interaction Model}

To test H2, I interact the policy treatment indicator with the pre-treatment log average degree, adapting the cross-sectional moderation framework of \citet{rajan1998}:
\begin{align}
\ln E[Y_{ct}] &= \beta \cdot \text{Post}_{ct} + \gamma \cdot \log(\bar{d}_c + 1) \nonumber \\
&\quad + \delta \cdot \text{Post}_{ct} \times \log(\bar{d}_c + 1) \nonumber \\
&\quad + \alpha_g + \lambda_t + \ln(\text{Total}_{ct})
\label{eq:interact}
\end{align}
where $\alpha_g$ denotes major-discipline fixed effects (which allow the time-invariant subcategory-level main effect $\gamma$ to be identified). The interaction coefficient $\delta$ captures whether denser pre-existing collaborative networks buffer communities against the policy shock. Standard errors in this specification are clustered at the subcategory level ($143$ clusters), aligning the clustering structure with the level of variation in the network moderator.

\subsection{Identification and Robustness Strategies}

To ensure that the baseline estimates reflect the impact of the endorsement policy rather than confounding secular trends or model misspecification, I implement three complementary identification strategies alongside small-sample inference diagnostics:

\paragraph{Dynamic Event Study.} 
To formally evaluate the parallel pre-trends assumption and trace the temporal dynamics of the treatment effect, I estimate a weekly leads-and-lags event study specification relative to the general policy enforcement date ($T_0 = \text{January 21, 2026}$) across all non-mathematics subcategories:
\begin{equation}
\ln\left(\frac{Y_{ct} + 0.5}{\text{Total}_{ct}}\right) = \sum_{k \neq -2} \beta_k \cdot \mathbb{1}[t = T_0 + k] + \alpha_c + \varepsilon_{ct}
\label{eq:eventstudy}
\end{equation}
where $\mathbb{1}[t = T_0 + k]$ denotes weekly event-time indicators (with $k = -2$ serving as the omitted reference week) and $\alpha_c$ absorbs time-invariant subcategory heterogeneity. Confirming that $\hat\beta_k \approx 0$ for $k < 0$ validates the absence of anticipatory pre-trends.

\paragraph{Difference-in-Differences (DiD) Design.} 
Because arXiv rolled out the tightened endorsement rule in mathematics six weeks prior to the rest of the platform, I exploit this staggered adoption window (October 1, 2025 to January 20, 2026) as a clean quasi-experiment~\cite{callaway2021difference}. Within this window, the 30 mathematics subcategories serve as the treated group while the remaining 107 non-mathematics subcategories serve as not-yet-treated controls:
\begin{equation}
\ln E[Y_{ct}] = \delta_{\text{DiD}} \cdot (\text{Math}_c \times \text{Post}^{\text{Dec}}_t) + \alpha_c + \gamma_t + \ln(\text{Total}_{ct})
\label{eq:did}
\end{equation}
where $\text{Math}_c$ indicates mathematics subcategories, $\text{Post}^{\text{Dec}}_t$ equals one after December 10, 2025, and fine-grained time fixed effects $\gamma_t$ purge common platform-wide shocks.

\paragraph{Daily Regression Discontinuity in Time (RDiT).} 
To isolate the immediate local impact of the policy right at the enforcement threshold, I construct a high-frequency daily panel within a $\pm 60$ day bandwidth around January 21, 2026 ($N = 121$ days) and estimate a sharp regression discontinuity model~\cite{cattaneo2020practical}:
\begin{equation}
\ln E[Y_{gd}] = \tau \cdot \text{Post}_d + f(R_d) + \text{Post}_d \times f(R_d) + \boldsymbol{\delta}_{\text{dow}} + \ln(\text{Total}_{gd})
\label{eq:rdd}
\end{equation}
where $R_d$ is the running variable measuring days relative to the January 21 cutoff, $f(R_d)$ is a flexible quadratic polynomial allowed to vary on either side of the threshold, and $\boldsymbol{\delta}_{\text{dow}}$ denotes day-of-week fixed effects capturing weekly submission cycles. Importantly, mathematics serves as an built-in falsification test at the January 21 cutoff because its treatment occurred six weeks earlier.

\paragraph{Small-Sample Cluster Diagnostics.} 
When clustering standard errors at the endorsement-domain level ($G = 8$), standard asymptotic sandwich estimators can be anti-conservative. In addition to applying \citet{cameron2008} $t(G-1) = t(7)$ degrees-of-freedom adjustments (critical values: $t_{0.05}(7) = 2.365$, $t_{0.01}(7) = 3.499$), I implement wild cluster bootstrap procedures~\cite{roodman2019fast} to verify that all primary inferences remain robust against small-cluster bias.

\subsection{Cross-Platform Spillover and Displacement Test}

A critical theoretical question is whether the sharp contraction in new submitters—particularly the nearly $50\%$ drop in quantitative biology (`q-bio`)—reflects a genuine decline in underlying scientific production or a behavioral shift in where researchers choose to disseminate their work (i.e., platform displacement). Because the Web hosts parallel open science infrastructure, researchers excluded by arXiv's social gatekeeping may migrate to alternative repositories with different governance models.

To disentangle platform displacement from secular disciplinary contraction, I design a cross-platform comparative test using \textbf{bioRxiv}, the premier preprint server for the biological sciences~\cite{sever2019biorxiv}, which experienced no contemporaneous change in its submission policies. I collect daily submission counts from bioRxiv over a 165-day window spanning the January 21, 2026 arXiv policy shock and estimate a Negative Binomial time-series model controlling for day-of-week seasonality ($\boldsymbol{\delta}_{\text{dow}}$) and a quadratic temporal trend ($f(R_d)$). If the collapse in arXiv's `q-bio` category were driven by a broader macro-level slowdown in biological research, bioRxiv should exhibit a parallel decline. Conversely, a stable or surging submission rate on bioRxiv post-January 21 would rule out field-wide research contraction and provide direct evidence of cross-platform spillover.

%==============================================================================
\section{Empirical Results}
%==============================================================================

\subsection{Main Policy Effect on New Entrants (H1)}

Table~\ref{tab:main} presents the estimated impact of tightening the endorsement policy on new-author entry. Across all model specifications, the policy triggered an immediate and substantial contraction in the influx of first-time submitters. In my preferred baseline specification (Column 1), the policy reduced the rate of new-author entry by $13.3\%$ ($\hat\beta = -0.142$, $p_{t(7)} = 0.029$). This negative effect remains highly robust across alternative specifications: replacing time fixed effects with subcategory-specific linear time trends (Column 2, $-17.8\%$), relaxing the unit-elasticity constraint by entering total submissions as a free covariate (Column 3, $-11.9\%$), and employing Negative Binomial estimation to explicitly parameterize overdispersion (Column 4, $-10.9\%$). Collectively, these estimates provide unambiguous empirical support for H1.

\begin{table}[t]
\centering
\begin{threeparttable}
\caption{Main Effect of Endorsement Policy on New Submitter Rate}
\label{tab:main}
\small
\setlength{\tabcolsep}{4pt}
\begin{tabular}{lcccc}
\toprule
& (1) & (2) & (3) & (4) \\
& Poisson & Poisson & Poisson & NB \\
\midrule
Post & $-0.142^{**}$ & $-0.196^{***}$ & $-0.127^{***}$ & $-0.115^{***}$ \\
& (0.052) & (0.013) & (0.038) & (0.034) \\
\addlinespace
$\log(\text{Total})$ & [offset] & [offset] & $0.905^{***}$ & [offset] \\
& & & (0.020) & \\
\midrule
Subcategory FE & Yes & Yes & Yes & No \\
Year-Month FE & Yes & --- & Yes & Yes \\
Subcat-specific trend & No & Yes & No & No \\
Major-category FE & --- & --- & --- & Yes \\
\midrule
Observations & 10,745 & 10,745 & 10,745 & 10,754 \\
Pseudo $R^2$ & 0.791 & 0.792 & 0.792 & 0.061 \\
Implied \% change & $-13.3\%$ & $-17.8\%$ & $-11.9\%$ & $-10.9\%$ \\
\bottomrule
\end{tabular}
\begin{tablenotes}
\small
\item \textit{Notes:} Standard errors are clustered at the endorsement-domain level (8 clusters). Statistical significance is evaluated using small-sample $t(7)$ critical values~\cite{cameron2008}: $p_{t(7)} = 0.029$ for Column (1). $^{***}p<0.001$, $^{**}p<0.01$, $^{*}p<0.05$.
\end{tablenotes}
\end{threeparttable}
\end{table}

\subsection{Network Moderation (H2)}

Table~\ref{tab:interact} reports the estimates from the network interaction model (Equation~\ref{eq:interact}). Consistent with H2, the interaction coefficient $\hat\delta$ is positive and statistically significant ($\hat\delta = +0.023$, $p = 0.005$ with subcategory clustering in Column 2; $p_{t(7)} = 0.098$ under conservative domain-level clustering in Column 1), showing that denser co-authorship networks buffer communities against the policy shock. At $\log(\bar{d}+1)=0$, the policy implies an $18.1\%$ reduction in newcomer entry ($\hat\beta = -0.200, p < 0.001$); each unit increase in network density offsets this penalty by about $2.3$ percentage points.

\begin{table}[t]
\centering
\begin{threeparttable}
\caption{Interaction Model: Moderation by Network Density}
\label{tab:interact}
\small
\setlength{\tabcolsep}{5pt}
\begin{tabular}{lcc}
\toprule
& (1) & (2) \\
& Domain cluster & Subcategory cluster \\
\midrule
Post & $-0.200^{**}$ & $-0.200^{***}$ \\
& (0.063) & (0.041) \\
\addlinespace
$\log(\bar{d} + 1)$ & $-0.016^{***}$ & $-0.016$ \\
& (0.003) & (0.022) \\
\addlinespace
Post $\times$ $\log(\bar{d} + 1)$ & $+0.023^{\dagger}$ & $+0.023^{**}$ \\
& (0.012) & (0.008) \\
\midrule
Major-category FE & Yes & Yes \\
Year-Month FE & Yes & Yes \\
Observations & 10,754 & 10,754 \\
Pseudo $R^2$ & 0.773 & 0.773 \\
\bottomrule
\end{tabular}
\begin{tablenotes}
\small
\item \textit{Notes:} Poisson QMLE models with $\log(\text{Total})$ entered as an exposure offset. Column (2) clusters standard errors at the subcategory level (143 clusters), matching the level of variation in the network moderator. Column (1) reports endorsement-domain clustering with $t(7)$ degrees-of-freedom adjustment ($p_{t(7)} = 0.098$). $^{***}p<0.001$, $^{**}p<0.01$, $^{\dagger}p<0.10$.
\end{tablenotes}
\end{threeparttable}
\end{table}

\subsection{Heterogeneous Disciplinary Exclusion (H3)}

To evaluate H3, I estimate the treatment effect separately across each major disciplinary group. As shown in Table~\ref{tab:hetero}, the endorsement tightening generated profound disciplinary stratification. In peripheral arXiv disciplines and fields characterized by sparse co-authorship ties, newcomer entry nearly collapsed: quantitative biology (`q-bio`) suffered a $48.9\%$ decline ($\hat\beta = -0.671, p = 0.017$), and economics (`econ`, $\bar{d} = 2.8$) experienced a $48.7\%$ drop ($\hat\beta = -0.668, p = 0.019$). Computer science (`cs`), which exhibits moderate network density ($\bar{d} = 13.9$) alongside a massive influx of interdisciplinary contributors, saw a statistically significant $21.4\%$ reduction ($\hat\beta = -0.241, p < 0.001$). By contrast, in densely connected fields such as physics ($\bar{d} = 89.6$) as well as core foundational arXiv domains such as mathematics ($\bar{d} = 2.4$), the policy effect was minimal and statistically indistinguishable from zero ($-6.1\%$ and $-6.2\%$, respectively). Together, these subsample dynamics corroborate H3: rather than acting as a uniform filter, the endorsement barrier induced near-exclusion for newcomers in sparse and peripheral scientific communities.

\begin{table*}[t]
\centering
\begin{threeparttable}
\caption{Heterogeneous Effects Across Academic Disciplines}
\label{tab:hetero}
\small
\begin{tabular}{lccccccc}
\toprule
& (1) & (2) & (3) & (4) & (5) & (6) & (7) \\
& q-bio & econ & cs & stat & math & physics & eess \\
\midrule
Post & $-0.671^{*}$ & $-0.668^{*}$ & $-0.241^{***}$ & $-0.122$ & $-0.064$ & $-0.063$ & $-0.052$ \\
& (0.281) & (0.286) & (0.049) & (0.146) & (0.105) & (0.055) & (0.150) \\
\midrule
Implied \% change & $-48.9\%$ & $-48.7\%$ & $-21.4\%$ & $-11.5\%$ & $-6.2\%$ & $-6.1\%$ & $-5.1\%$ \\
Avg Degree ($\bar{d}$) & 14.1 & 2.8 & 13.9 & 3.8 & 2.4 & 89.6 & 8.6 \\
Observations & 398 & 276 & 2,829 & 362 & 2,595 & 3,747 & 376 \\
Pseudo $R^2$ & 0.217 & 0.213 & 0.878 & 0.435 & 0.453 & 0.572 & 0.437 \\
\bottomrule
\end{tabular}
\begin{tablenotes}
\small
\item \textit{Notes:} Each column reports a separate Poisson QMLE regression including subcategory fixed effects, year-month fixed effects, and the log total submission exposure offset. Standard errors are clustered at the subcategory level. Disciplines are ordered from left to right by the magnitude of the negative treatment effect. $^{***}p<0.001$, $^{*}p<0.05$.
\end{tablenotes}
\end{threeparttable}
\end{table*}

\subsection{Validity and Identification}

\subsubsection{Dynamic Event Study}
\begin{figure}[t]
\centering
\includegraphics[width=\columnwidth]{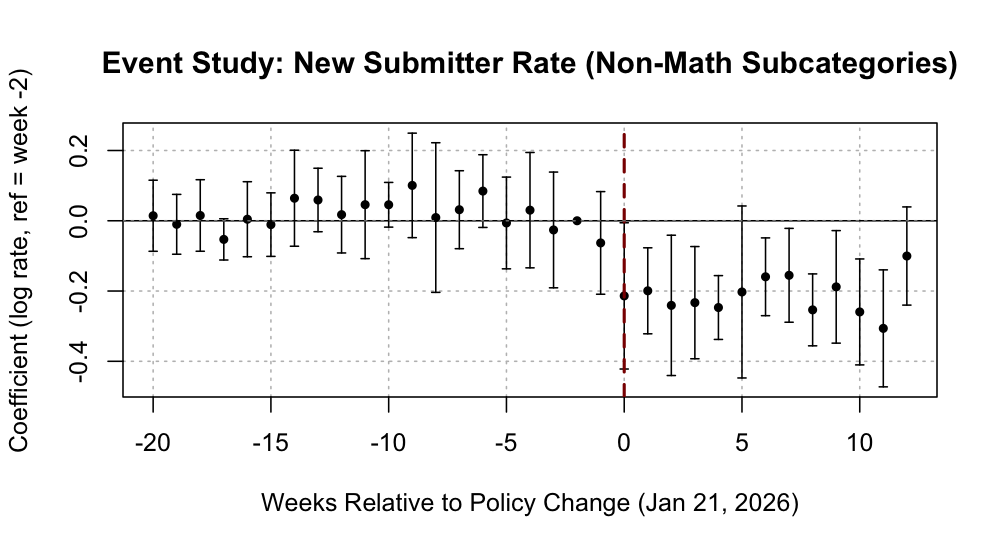}
\caption{Weekly event study coefficients relative to policy onset (Jan 21, 2026) for non-math subcategories. Error bars show 95\% confidence intervals. Pre-trends centered around zero; post-policy decline is immediate and sustained.}
\Description{Event study plot showing weekly regression coefficients before and after policy onset.}
\label{fig:es}
\end{figure}

To evaluate the parallel pre-trends assumption, we estimate weekly leads and lags relative to the general policy date ($T_0 = \text{Jan 21, 2026}$) for all non-mathematics subcategories:
\begin{equation}
\ln\left(\frac{Y_{ct} + 0.5}{\text{Total}_{ct}}\right) = \sum_{k \neq -2} \beta_k \cdot \mathbb{1}[t = T_0 + k] + \alpha_c + \varepsilon_{ct}
\end{equation}
Figure~\ref{fig:es} plots the estimated weekly coefficients. Pre-treatment coefficients fluctuate closely around zero (mean $= 0.02$, maximum $|\hat\beta_k| = 0.10$), confirming parallel trends prior to intervention. Immediately upon policy implementation, coefficients drop sharply into negative territory (mean post-treatment $\hat\beta_k = -0.21$) and remain persistently depressed throughout the post-treatment window.

\subsubsection{Sharp Discontinuity at Policy Cutoff}
\begin{figure}[t]
\centering
\includegraphics[width=\columnwidth]{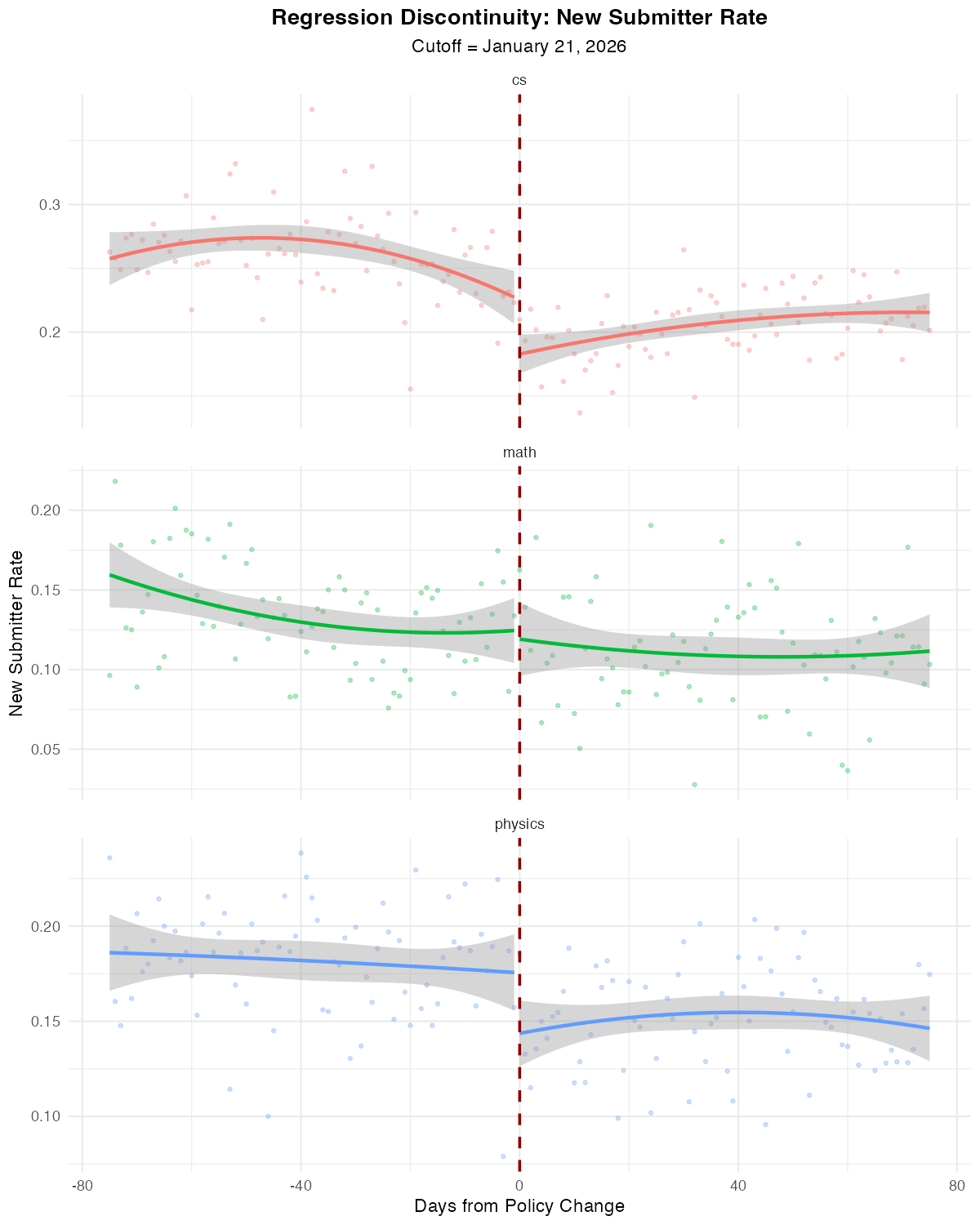}
\caption{Regression discontinuity visualization ($\pm 75$ days around Jan 21, 2026). Computer science and physics exhibit sharp negative discontinuities; mathematics exhibits no break at this cutoff due to early December treatment.}
\Description{Regression discontinuity plot showing daily new submitter rates around the January 21 cutoff.}
\label{fig:rdd}
\end{figure}

Using daily data within a $\pm 60$ day window around January 21, 2026, we estimate daily regression discontinuity models controlling for day-of-week fixed effects and quadratic running variables~\cite{cattaneo2020practical}. As plotted in Figure~\ref{fig:rdd}, computer science exhibits an immediate $-20.0\%$ discontinuity ($\hat\tau = -0.223, p < 0.001$) and physics an immediate $-18.6\%$ discontinuity ($\hat\tau = -0.206, p = 0.025$). Mathematics shows no discontinuity at January 21 ($\hat\tau = -0.108, p = 0.456$) because it was pre-treated six weeks earlier.

\subsection{Identification and Robustness Checks}

The complementary checks described in Section~3.3 support the baseline interpretation. In Figure~\ref{fig:es}, event-study coefficients remain near zero before the January 21, 2026 policy onset and fall sharply afterward, indicating parallel pre-trends and a persistent entry decline. The staggered DiD design yields a $6.1\%$ decline for mathematics after its December rollout ($\hat\delta_{\text{DiD}} = -0.063, p_{t(7)} = 0.0006$), while the RDiT estimates in Figure~\ref{fig:rdd} show sharp drops for computer science and physics but no break for mathematics at the January cutoff, consistent with its earlier treatment.

For inference, the baseline treatment coefficient ($\hat\beta = -0.142$) gives $p_{t(7)} = 0.029$ under eight-domain clustering, and subcategory clustering in the interaction model gives $p = 0.005$. Wild cluster bootstrap checks~\cite{cameron2008, roodman2019fast} confirm that the primary conclusions are robust to small-cluster bias.

\subsection{Mechanism Analysis: Platform Displacement vs. Research Contraction}

Having established that arXiv's endorsement tightening severely curtailed newcomer entry in sparsely networked and peripheral fields, I now examine the behavioral mechanism underlying this decline: did excluded researchers cease producing preprints altogether, or did the barrier trigger cross-platform displacement toward alternative open science venues?

\begin{figure}[t]
\centering
\includegraphics[width=0.88\columnwidth]{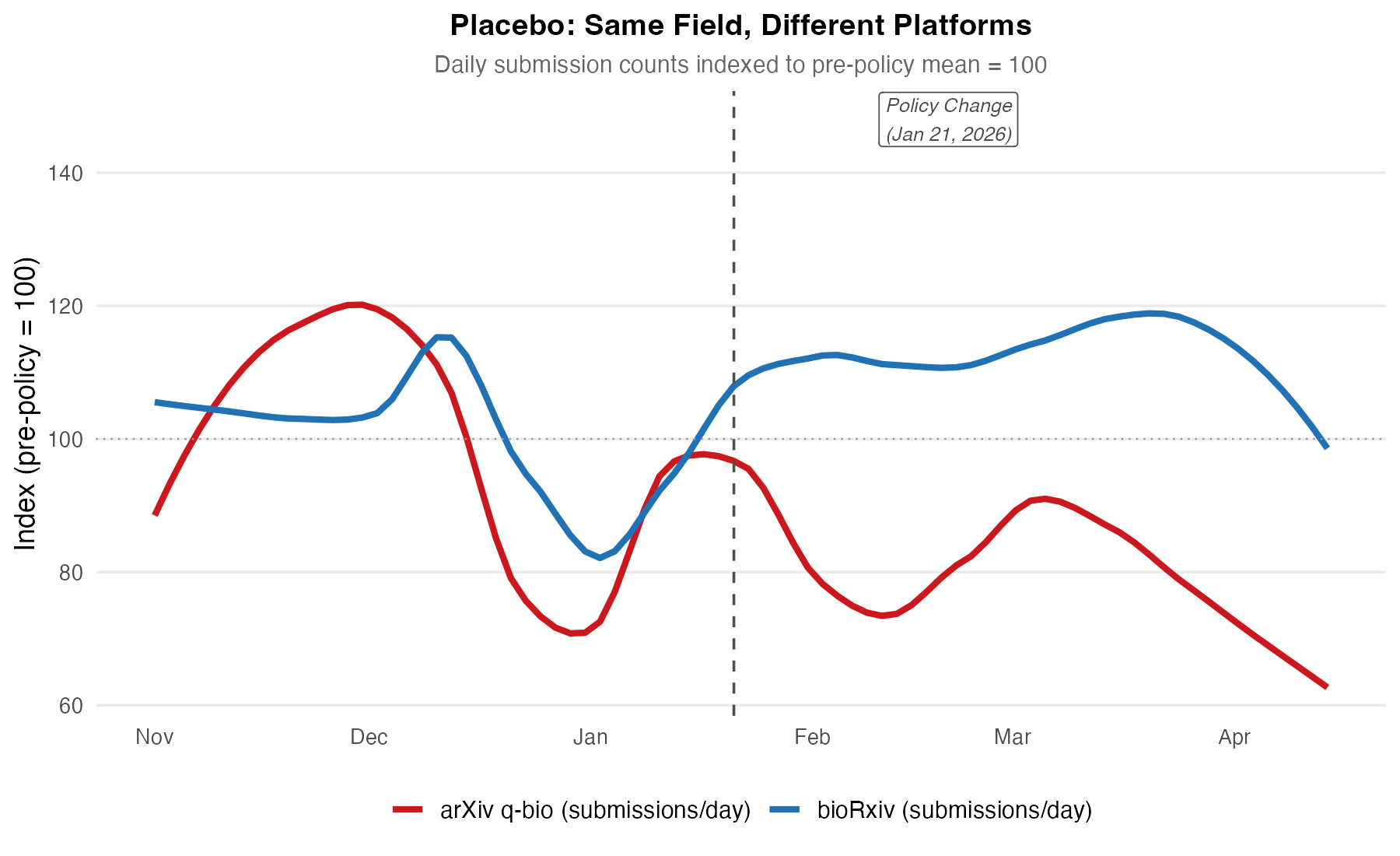}
\caption{Indexed daily activity (pre-policy mean = 100): bioRxiv submissions (blue) remained stable and trended upward after January 21, 2026, whereas arXiv quantitative biology new submitters (red) dropped precipitously at the policy cutoff.}
\Description{Placebo comparison line plot comparing bioRxiv daily submissions to arXiv quantitative biology submitters.}
\label{fig:placebo}
\end{figure}

Figure~\ref{fig:placebo} presents the cross-platform comparison between arXiv's quantitative biology (`q-bio`) category and bioRxiv. ArXiv `q-bio` new submitters fell from a pre-policy average of $3.3$ per day to $1.5$ after January 21, 2026, a raw contraction of $53.1\%$ that mirrors the $-48.9\%$ Poisson QMLE estimate in Table~\ref{tab:hetero}. In contrast, bioRxiv submissions increased by $39.4\%$ ($\hat\phi = +0.332, p = 0.0095$), rising from a pre-policy daily mean of $178.2$ papers to $202.1$ papers.

Figure~\ref{fig:placebo} visualizes these diverging trajectories by indexing both series to a pre-policy baseline of 100. While arXiv `q-bio` new entrants (red curve) suffered a precipitous structural break immediately after the January 21 cutoff, bioRxiv activity (blue curve) maintained a robust upward trajectory. This sharp divergence provides two critical insights. First, it decisively rules out the alternative explanation that the collapse in arXiv `q-bio` was caused by an exogenous, field-wide slowdown in biological research during early 2026. Second, it demonstrates that network-mediated gatekeeping on a dominant Web platform induces a \textbf{platform displacement effect}: when newcomer biologists—who typically lack dense co-authorship ties within arXiv's physics- and math-dominated ecosystem—encountered prohibitive social endorsement barriers on arXiv, they redirected their scientific outputs toward bioRxiv, where access relies on content screening rather than pre-existing interpersonal networks.

%==============================================================================
\section{Discussion}
%==============================================================================

In this study, I leverage arXiv's tightening of its endorsement policy as a quasi-experiment to investigate how entry barriers in Web-based scientific production disproportionately affect contributor participation across disciplines. The findings indicate that although the endorsement policy was introduced to curb the influx of low-quality submissions, it generated a profound unintended consequence: by tethering platform access to interpersonal vouching, the policy produced new forms of structural inequality contingent on the underlying network density of different scientific communities.

\subsection{Theoretical Implications: Bridging Platform Governance and Academic Stratification}

This study makes two primary theoretical contributions. First, by shifting the analytical lens to the ``very point of entry,'' this work extends classical theories of the Matthew Effect and cumulative advantage in science~\cite{merton1968, clauset2015systematic, way2019productivity}. Prior research on inequality in scientific production has predominantly focused on disparities that unfold after researchers have already entered the scholarly arena---such as institutional prestige biases in peer review~\cite{teplitskiy2022status}, unequal funding allocation~\cite{ginther2011race}, and skewed citation accumulation~\cite{dong2017century}---while implicitly treating the initial opportunity to participate as universally accessible. Even studies celebrating preprint servers for democratizing scholarly communication often assume that once a platform is publicly available on the Web, the playing field is leveled~\cite{bjork2010open, xie2021preprint}. By contrast, this study demonstrates that stratification mechanisms begin operating before a paper is even read or evaluated. When open science infrastructure relies on network-mediated gatekeeping, the co-authorship network topologies documented by \citet{newman2001} and \citet{lariviere2013bibliometrics} cease to be mere conduits for collaboration and norm transmission; instead, they evolve into the arbiters of who is permitted to contribute to the scientific record. In doing so, this study shows that the Matthew Effect not only governs the distribution of recognition and rewards among active scientists, but also actively polices the boundary of scientific participation itself.

Second, this work advances the literature in Web science and platform governance by reconceptualizing how platform rules interact with community social structures. Existing studies on online gatekeeping, credentialing, and content moderation typically model entry barriers as uniform, platform-wide attributes that impose constant friction costs across all prospective newcomers~\cite{kraut2012building, halfaker2013rise, kumar2015analysis, jhaver2019transparency}. By integrating network theories of structural holes and tie density, I show that a single, formally neutral governance rule binds with vastly different severity depending on the relational topology of the subcommunity to which it is applied. In densely connected disciplines, abundant collaborative ties absorb interpersonal vouching requirements with minimal friction; in fragmented or peripheral disciplines, the exact same rule demands substantial social capital~\cite{lin2001}, triggering severe exclusion and cross-platform displacement. This insight moves platform governance theory beyond treating communities as homogeneous user bases, demonstrating that the fairness of an online rule cannot be evaluated in isolation from the offline social networks in which its users are embedded.

\subsection{Practical Implications: When Quality Control Becomes an Engine of Inequality}

From a practical standpoint, this study highlights a critical governance paradox: a well-intentioned policy instrument designed to safeguard content quality ultimately mutated into a mechanism of structural exclusion. The results uncover a subtle failure dynamic within scholarly infrastructure: because arXiv delegated quality screening to interpersonal endorsement, the policy did not filter submissions based on their intrinsic scholarly rigor, but rather screened entrants based on their pre-existing social connectivity~\cite{nilizadeh2019think, teplitskiy2022status}. Consequently, the burden of quality control did not fall on spammers in high-volume core fields, but on legitimate scholars working in sparsely networked or emerging interdisciplinary domains.

For platform architects and open science stewards, this study offers several actionable design implications, such as decoupling quality control from interpersonal social capital, or—where human moderation remains necessary—relying on centralized or rotating editorial screening pools rather than forcing newcomers to privately solicit personal favors from established insiders.

\subsection{Future Research Directions}

Future work can move beyond examining participation volume to further trace the epistemic trajectories of displaced knowledge. By combining full-text semantic embeddings with cross-platform author matching across arXiv, bioRxiv, SSRN, and downstream peer-reviewed journals, researchers can investigate whether manuscripts redirected by platform gatekeeping experience delays in dissemination, altered citation velocities, or shifts in the tiers of their eventual publication venues~\cite{wang2013quantifying, fortunato2018science}.

Alternatively, future studies can extend the macro-level subcategory analysis presented here to the micro-level dynamics of individual scholars' ego-networks. Constructing longitudinal co-authorship networks disambiguated at the individual author level will enable researchers to track how newcomers actively navigate and adapt to new institutional barriers.

%==============================================================================
\section{Conclusion}
%==============================================================================

Open science platforms are built on the foundational promise of lowering barriers to scholarly participation, yet maintaining order within an open commons requires ongoing governance. By examining arXiv's tightening of its endorsement policy in 2026, this study demonstrates that when platforms rely on existing social networks to police entry boundaries, quality control comes at the expense of structural equity. While the policy substantially curtailed the influx of new authors, its burden fell overwhelmingly on academic communities with sparse collaborative ties—triggering cross-platform displacement and reinforcing historical disciplinary hierarchies. Ultimately, ensuring that the Web remains an inclusive engine for scientific discovery requires designing governance architectures that evaluate what newcomers contribute, rather than who they know.

%==============================================================================
% Bibliography
%==============================================================================
\bibliographystyle{ACM-Reference-Format}
\bibliography{references}

\end{document}